\documentclass[%
a4paper,
prd,
twocolumn,
superscriptaddress,
preprintnumbers,
nofootinbib,
nobibnotes,
amsmath,amssymb,
aps,
fleqn,
floatfix
]{revtex4-2}
\usepackage{amsfonts,graphics,color}
\usepackage[english]{babel}
\usepackage{graphicx}
\usepackage{amsmath}
\usepackage{dcolumn}
\usepackage{bm}
\usepackage{multirow}
\usepackage{xcolor}
\definecolor{xlinkcolor}{cmyk}{1,1,0,0}
\usepackage[bookmarks=true, pdfnewwindow=true, colorlinks=true, linkcolor=xlinkcolor, citecolor=xlinkcolor, filecolor=xlinkcolor, urlcolor=xlinkcolor, final=true]{hyperref}

\usepackage[normalem]{ulem}
\definecolor{myblue}{rgb}{0.05,0.1,0.5}

\begin{document}

\preprint{INR-TH-2025-001}

\title{Ultra-high-energy event KM3-230213A constraints on Lorentz Invariance Violation in neutrino sector}

\author{Petr Satunin}
\affiliation{Institute for Nuclear
Research of the Russian Academy of Sciences, 60th October Anniversary Prospect 7a, Moscow 117312, Russia}

\begin{abstract}
  We discuss the constraints on superluminal neutrino Lorentz Invariance Violation (LIV) parameters from the observation of the ultra-high-energy event  KM3-230213A   by the KM3NeT collaboration in cases of linear $n=1$ and quadratic $n=2$ LIV scenarios. Assuming extragalactic origin of the event, we obtain the constraints on LIV mass scale $\Lambda_{n=1} =  1.1 \times 10^{30}\, \mbox{GeV}$ and $\Lambda_{n=2} =  1.1 \times 10^{19}\, \mbox{GeV}$  from the absence of neutrino splitting.   
\end{abstract}
\maketitle

\paragraph{Introduction} Recently, KM3NeT collaboration reported \cite{Aiello2025} the  event called KM3-230213A that has been identified with  astrophysical  neutrino of energy $220^{+570}_{-100}$ PeV. This energy is almost two orders of magnitude larger than the previous most energetic neutrino event detected at IceCube. The origin of the neutrino event KM3-230213A is an intriguing question: galactic \cite{Adriani:2025mib}, extragalactic \cite{KM3NeT:2025bxl} and cosmogenic \cite{KM3NeT:2025vut} possible origins are discussed. Such ultra high energy of the event allows us to test some fundamental principles such as Lorentz Invariance  at an unprecedented level. Small departures of LI appear in some models of quantum gravity; see \cite{Addazi:2021xuf} for a review. 

The most sensitive to high energies models of Lorentz Invariance Violation (LIV) are characterized by the class of dispersion relations 
\begin{equation}
\label{eq:DispRelGeneral}
    E^2=m^2+k^2\left(1+s_n\left(\frac{k}{\Lambda}\right)^n\right),
\end{equation}
where $n=1$ and $n=2$ respond to linear and quadratic LIV correspondingly; $\Lambda$ is the energy scale at which LI is broken. $s_n$ is the sign which strongly influences the LIV phenomenology: $s_n = +1\, (-1)$ called superluminal (subluminal) LIV which lead to superluminal (subluminal) velocity of a given particle. Besides, LIV leads to the modification of thresholds and cross-sections of particle processes; superluminal LIV in general leads to several decay channels of a given particle absent in the LI case; subluminal LIV, in contrast, suppresses some interactions present in LI. 

In this letter we show that the detection of the neutrino event leads to neutrino (meta)stability (or long lifetime) which in turn leads to the absence of concrete decay channels. Consequently,  we limit ourselves to consider {\it superluminal} neutrino. The corresponding decay channels are neutrino splitting 
 $\nu \to \nu \nu\bar{\nu}$,  discussed in \cite{Stecker:2014oxa, Jentschura:2020nfe, Carmona:2022dtp} and neutrino  pair production $\nu \to \nu e^+e^-$ \cite{Cohen:2011hx,Huo:2011ve,Stecker:2014oxa}. Before the numerical consideration of the processes, let us provide the best current bound\footnote{Originally $c_6 > -5.2 \cdot 10^{-35}\ \mbox{GeV}^{-2}$ in terms of the Data Tables \cite{Kostelecky:2008ts}.} on superluminal LIV from IceCube  \cite{Stecker:2014oxa},
\begin{equation}
    \Lambda = 1.4 \times 10^{17}\,\mbox{GeV}, \qquad n=2.
\end{equation}

\paragraph{EFT Lagrangian}

In order to consider cross-sections of neutrino-involved processes, one should start from the effective field theory (EFT)  Lagrangian. 
The minimal LIV addition to the Standard Model (SM) Lagrangian, preserving $SU(2)_L$ gauge symmetry reads,
\begin{equation}
    {\cal L}_\text{LIV}^{(\nu), n} =  \frac{1}{\Lambda^n} \, \bar{l^\alpha}_{L} \gamma^0 \, (iD_0)^{n+1} \, l^{\alpha}_{L}\,, 
\label{eq:Lagr}
\end{equation}
where $l_L^\alpha = (\nu_\alpha,l_\alpha)_L$ is the left lepton doublet, $\alpha$ counts the lepton generations, $D_0$ is the time part of the SM covariant derivative. Note that $n=1$ scenario is CPT-odd while $n=2$ is CPT-even.  Considering only the quadratic over neutrino part of  (\ref{eq:Lagr}), one obtains the dispersion relation, 
\begin{align}
   E^2 = k^2 \pm \frac{k^3}{\Lambda},  \qquad n=1, \\
     E^2 = k^2 + \frac{k^4}{\Lambda^2},  \qquad n=2.
\end{align}
Note that for $n=1$ the sign before the LIV term is different for neutrino and antineutrino. Although the Lagrangian (\ref{eq:Lagr}) includes modified interactions, the main LIV effect on decay widths is due to modified kinematics. Next, we discuss the processes.

\paragraph{Neutrino splitting $\nu \to \nu\nu\bar{\nu}$} In order to take a numerical value of the neutrino width associated with neutrino splitting, we follow the analytical result of Carmona et al \cite{Carmona:2022dtp} obtained in the collinear approximation neglecting neutrino masses. The width of the neutrino splitting does not depend on flavour and reads \cite{Carmona:2022dtp},
\begin{align}
   \Gamma_{\nu\to \nu\nu\bar{\nu}}\approx 
 G_F^2  \,
\frac{E^5}{6\,\pi^3}  \left(\frac{E}{\Lambda}\right)^{3n} c^{(\nu)} .
\label{eq:Gamma-nu-nubar}
\end{align}
Here $c^{(\nu)} \approx 0.024$, $G_F$ is the Fermi constant. More detailed characteristics are the neutrino energy losses, see \cite{Carmona:2022dtp}. We leave this approach for a future more detailed analysis. 
Solving eq.~(\ref{eq:Gamma-nu-nubar}) with respect to $\Lambda$, one obtains 
\begin{equation}
\label{eq:Lambda_split}
    \Lambda = \begin{cases}
        \left(1.3 \cdot 10^{-4}\, G_F^2 L E^5\right)^{1/3}\times E, & n=1, \\ 
        \left( 1.3 \cdot 10^{-4}\, G_F^2 L E^{5}\right)^{1/6}\times E, & n=2,
    \end{cases}
\end{equation}
where $L =   \Gamma_{\nu\to \nu\nu\bar{\nu}}^{-1}$ is the mean free path of the neutrino associated with the splitting. We take it as the distance to the possible source.

Further we use the median value of KM3-230213A energy with $68 \%$ CL errorbars, $E=220^{+570}_{-100}$ PeV, and 
consider three scenarios of neutrino origin: galactic, extragalactic and cosmogenic ones. In the galactic scenario, we take $L = 10$ kpc, so eq.~(\ref{eq:Lambda_split}) reads,
\begin{equation}
\label{galactic}
    \Lambda = \begin{cases}
       5.3^{+156.2}_{-4.2} \times 10^{29}\, \mbox{GeV}, & n=1, \\ 
       1.1_{-0.7}^{+10.2} \times 10^{19}\, \mbox{GeV}, & n=2.
    \end{cases}
\end{equation}
Assuming extragalactic origin of neutrino\footnote{cf. other ultra high energy events came from that distance \cite{Kuznetsov:2023jfw}.} we can take $L = 10$ Mpc, obtaining
\begin{equation}
\label{extragalactic}
    \Lambda = \begin{cases}
       5.3^{+156.2}_{-4.2} \times 10^{30}\, \mbox{GeV}, & n=1, \\ 
       3.4^{+32.3}_{-2.3} \times 10^{19}\, \mbox{GeV}, & n=2.
    \end{cases}
\end{equation}
Assuming cosmogenic origin we can put $L = 100$ Mpc, obtaining
\begin{equation}
\label{cosmogenic}
    \Lambda = \begin{cases}
       1.2^{+33.6}_{-1.0}  \times 10^{31}\, \mbox{GeV}, & n=1, \\ 
       5.0^{+47.4}_{-3.3}  \times 10^{19}\, \mbox{GeV}, & n=2.
    \end{cases}
\end{equation}

Splitting cannot produce significant excess at lower energy in a spectrum decreasing with energy like $\sim E^{-2}$. In fact, splitting of a neutrino of energy $E$ into 3 particles of energy $E/3$ adds 3 particles to the spectrum which consists of $(E/(E/3))^2$ = 9 particles.

\paragraph{Neutrino $e^+e^-$ pair production}

For the aforementioned values of $\Lambda$ and $E$ one cannot neglect the electron mass values, so one should estimate the threshold effect, which is in general complicated for non-equal 3 particles in the final state. Consider for simplicity collinear kinematics, neutrino of energy and x-projection of momentum $(k+\frac{k^3}{2\Lambda^2},k)$ decays to a neutrino, an electron and a positron each of which takes $k/3$ momentum. Finally, the threshold of the effective decay reads,
\begin{equation}
    \Lambda_{thr} = \frac{E^2}{\sqrt{6}m_e} = 3.9_{-2.7}^{+46.0} \times 10^{19}\,\mbox{GeV}, \qquad  n=2.
\end{equation}
For larger $\Lambda$ the process does not occur, for lower $\Lambda$ the calculation is similar to the neutrino splitting. Carmona et al \cite{Carmona:2022dtp} calculated the corresponding width in case of zero electron mass; the width is that of eq.~(\ref{eq:Lambda_split}) multiplied by $0.13$. Finally, for our values of parameters $\nu\to\nu e^+e^-$ is subleading with respect to neutrino splitting.  

\paragraph{LIV constraints from the muon neutrino production in pion decay}

Consider the opposite process, the production of muon neutrino. 
In the majority of models, neutrino is produced very fast in the process of pion decay, $\pi^+ \to \mu^+ \nu_\mu$.  If the process is kinematically forbidden, no neutrino is produced. Two-particle decay threshold rule reads,
\begin{equation}
    m^2_{eff, \nu} \equiv \frac{E^4}{\Lambda^2} = m_\pi^2 - m_\mu^2 = \left(91 \, \mbox{MeV}\right)^2, \qquad n=2.
\end{equation}
From that we obtain a constraint,
\begin{equation}
    \Lambda = 5.3_{-3.7}^{+5.8} \times 10^{17} \, \mbox{GeV}, \qquad n=2.
\end{equation}
It is an order of magnitude weaker than the splitting ones, (\ref{galactic})-(\ref{cosmogenic})
\paragraph{Conclusion} We estimated constraints on LIV in the neutrino sector from the event KM3-230213A. It turns out that the best constraints came from the absence of the neutrino splitting.  We considered separately the cases for neutrino of galactic, extragalactic, and cosmogenic origin. Since it is concluded in \cite{Adriani:2025mib} that the galactic origin is unlikely, we may take  the lower bound of the 
constraint interval (\ref{extragalactic}) as the conservative $68 \%$ CL bound, precisely
\begin{equation}
    \Lambda = \begin{cases}
       1.1 \times 10^{30}\, \mbox{GeV}, & n=1, \\ 
       1.1 \times 10^{19}\, \mbox{GeV}, & n=2.
    \end{cases}
\end{equation}
In order to obtain more precise statistically significant constraints, the models of galactic, extragalactic, or cosmogenic sources are required, as well as an accurate calculation of neutrino energy losses.

After the appearance of the first version of this manuscript, LIV analysis of the  KM3-230213A event  by the KM3NET collaboration appears \cite{KM3NeT:2025mfl}. This analysis is related to the scenario of constant superluminal neutrino velocity, or $n=0$ in our notations, see eq.(\ref{eq:DispRelGeneral}). In this scenario, the neutrino splitting does not occur due to the energy-momentum conservation while the pair production $\nu\to \nu e^+e^-$ occurs over a certain threshold in the case of standard electrons.

\paragraph{Acknowledgements} The author thanks Grigory Rubtsov and Mikhail Kuznetsov for prompt discussions.

\bibliography{bibl}

\begin{thebibliography}{13}%
\makeatletter
\providecommand \@ifxundefined [1]{%
 \@ifx{#1\undefined}
}%
\providecommand \@ifnum [1]{%
 \ifnum #1\expandafter \@firstoftwo
 \else \expandafter \@secondoftwo
 \fi
}%
\providecommand \@ifx [1]{%
 \ifx #1\expandafter \@firstoftwo
 \else \expandafter \@secondoftwo
 \fi
}%
\providecommand \natexlab [1]{#1}%
\providecommand \enquote  [1]{``#1''}%
\providecommand \bibnamefont  [1]{#1}%
\providecommand \bibfnamefont [1]{#1}%
\providecommand \citenamefont [1]{#1}%
\providecommand \href@noop [0]{\@secondoftwo}%
\providecommand \href [0]{\begingroup \@sanitize@url \@href}%
\providecommand \@href[1]{\@@startlink{#1}\@@href}%
\providecommand \@@href[1]{\endgroup#1\@@endlink}%
\providecommand \@sanitize@url [0]{\catcode `\\12\catcode `\$12\catcode `\&12\catcode `\#12\catcode `\^12\catcode `\_12\catcode `\%12\relax}%
\providecommand \@@startlink[1]{}%
\providecommand \@@endlink[0]{}%
\providecommand \url  [0]{\begingroup\@sanitize@url \@url }%
\providecommand \@url [1]{\endgroup\@href {#1}{\urlprefix }}%
\providecommand \urlprefix  [0]{URL }%
\providecommand \Eprint [0]{\href }%
\providecommand \doibase [0]{https://doi.org/}%
\providecommand \selectlanguage [0]{\@gobble}%
\providecommand \bibinfo  [0]{\@secondoftwo}%
\providecommand \bibfield  [0]{\@secondoftwo}%
\providecommand \translation [1]{[#1]}%
\providecommand \BibitemOpen [0]{}%
\providecommand \bibitemStop [0]{}%
\providecommand \bibitemNoStop [0]{.\EOS\space}%
\providecommand \EOS [0]{\spacefactor3000\relax}%
\providecommand \BibitemShut  [1]{\csname bibitem#1\endcsname}%
\let\auto@bib@innerbib\@empty
\bibitem [{\citenamefont {Aiello~et al}(2025)}]{Aiello2025}%
  \BibitemOpen
  \bibfield  {author} {\bibinfo {author} {\bibfnamefont {S.}~\bibnamefont {Aiello~et al}},\ }\bibfield  {title} {\bibinfo {title} {Observation of an ultra-high-energy cosmic neutrino with km3net},\ }\href {https://doi.org/10.1038/s41586-024-08543-1} {\bibfield  {journal} {\bibinfo  {journal} {Nature}\ }\textbf {\bibinfo {volume} {638}},\ \bibinfo {pages} {376} (\bibinfo {year} {2025})}\BibitemShut {NoStop}%
\bibitem [{\citenamefont {Adriani}\ \emph {et~al.}(2025{\natexlab{a}})\citenamefont {Adriani} \emph {et~al.}}]{Adriani:2025mib}%
  \BibitemOpen
  \bibfield  {author} {\bibinfo {author} {\bibfnamefont {O.}~\bibnamefont {Adriani}} \emph {et~al.},\ }\bibfield  {title} {\bibinfo {title} {{On the Potential Galactic Origin of the Ultra-High-Energy Event KM3-230213A}},\ }\href@noop {} {\  (\bibinfo {year} {2025}{\natexlab{a}})},\ \Eprint {https://arxiv.org/abs/2502.08387} {arXiv:2502.08387 [astro-ph.HE]} \BibitemShut {NoStop}%
\bibitem [{\citenamefont {Baldini}\ \emph {et~al.}(2025)\citenamefont {Baldini} \emph {et~al.}}]{KM3NeT:2025bxl}%
  \BibitemOpen
  \bibfield  {author} {\bibinfo {author} {\bibfnamefont {P.}~\bibnamefont {Baldini}} \emph {et~al.} (\bibinfo {collaboration} {KM3NeT, MessMapp Group, Fermi-LAT, Owens Valley Radio Observatory 40-m Telescope Group, SVOM}),\ }\bibfield  {title} {\bibinfo {title} {{Characterising Candidate Blazar Counterparts of the Ultra-High-Energy Event KM3-230213A}},\ }\href@noop {} {\  (\bibinfo {year} {2025})},\ \Eprint {https://arxiv.org/abs/2502.08484} {arXiv:2502.08484 [astro-ph.HE]} \BibitemShut {NoStop}%
\bibitem [{\citenamefont {Adriani}\ \emph {et~al.}(2025{\natexlab{b}})\citenamefont {Adriani} \emph {et~al.}}]{KM3NeT:2025vut}%
  \BibitemOpen
  \bibfield  {author} {\bibinfo {author} {\bibfnamefont {O.}~\bibnamefont {Adriani}} \emph {et~al.} (\bibinfo {collaboration} {KM3NeT}),\ }\bibfield  {title} {\bibinfo {title} {{On the potential cosmogenic origin of the ultra-high-energy event KM3-230213A}},\ }\href@noop {} {\  (\bibinfo {year} {2025}{\natexlab{b}})},\ \Eprint {https://arxiv.org/abs/2502.08508} {arXiv:2502.08508 [astro-ph.HE]} \BibitemShut {NoStop}%
\bibitem [{\citenamefont {Addazi}\ \emph {et~al.}(2022)\citenamefont {Addazi} \emph {et~al.}}]{Addazi:2021xuf}%
  \BibitemOpen
  \bibfield  {author} {\bibinfo {author} {\bibfnamefont {A.}~\bibnamefont {Addazi}} \emph {et~al.},\ }\bibfield  {title} {\bibinfo {title} {{Quantum gravity phenomenology at the dawn of the multi-messenger era\textemdash{}A review}},\ }\href {https://doi.org/10.1016/j.ppnp.2022.103948} {\bibfield  {journal} {\bibinfo  {journal} {Prog. Part. Nucl. Phys.}\ }\textbf {\bibinfo {volume} {125}},\ \bibinfo {pages} {103948} (\bibinfo {year} {2022})},\ \Eprint {https://arxiv.org/abs/2111.05659} {arXiv:2111.05659 [hep-ph]} \BibitemShut {NoStop}%
\bibitem [{\citenamefont {Stecker}\ \emph {et~al.}(2015)\citenamefont {Stecker}, \citenamefont {Scully}, \citenamefont {Liberati},\ and\ \citenamefont {Mattingly}}]{Stecker:2014oxa}%
  \BibitemOpen
  \bibfield  {author} {\bibinfo {author} {\bibfnamefont {F.~W.}\ \bibnamefont {Stecker}}, \bibinfo {author} {\bibfnamefont {S.~T.}\ \bibnamefont {Scully}}, \bibinfo {author} {\bibfnamefont {S.}~\bibnamefont {Liberati}},\ and\ \bibinfo {author} {\bibfnamefont {D.}~\bibnamefont {Mattingly}},\ }\bibfield  {title} {\bibinfo {title} {{Searching for Traces of Planck-Scale Physics with High Energy Neutrinos}},\ }\href {https://doi.org/10.1103/PhysRevD.91.045009} {\bibfield  {journal} {\bibinfo  {journal} {Phys. Rev. D}\ }\textbf {\bibinfo {volume} {91}},\ \bibinfo {pages} {045009} (\bibinfo {year} {2015})},\ \Eprint {https://arxiv.org/abs/1411.5889} {arXiv:1411.5889 [hep-ph]} \BibitemShut {NoStop}%
\bibitem [{\citenamefont {Jentschura}(2020)}]{Jentschura:2020nfe}%
  \BibitemOpen
  \bibfield  {author} {\bibinfo {author} {\bibfnamefont {U.~D.}\ \bibnamefont {Jentschura}},\ }\bibfield  {title} {\bibinfo {title} {{Squeezing the Parameter Space for Lorentz Violation in the Neutrino Sector with Additional Decay Channels}},\ }\href {https://doi.org/10.3390/particles3030041} {\bibfield  {journal} {\bibinfo  {journal} {Particles}\ }\textbf {\bibinfo {volume} {3}},\ \bibinfo {pages} {630} (\bibinfo {year} {2020})},\ \Eprint {https://arxiv.org/abs/2009.11947} {arXiv:2009.11947 [hep-ph]} \BibitemShut {NoStop}%
\bibitem [{\citenamefont {Carmona}\ \emph {et~al.}(2023)\citenamefont {Carmona}, \citenamefont {Cort\'es}, \citenamefont {Relancio},\ and\ \citenamefont {Reyes}}]{Carmona:2022dtp}%
  \BibitemOpen
  \bibfield  {author} {\bibinfo {author} {\bibfnamefont {J.~M.}\ \bibnamefont {Carmona}}, \bibinfo {author} {\bibfnamefont {J.~L.}\ \bibnamefont {Cort\'es}}, \bibinfo {author} {\bibfnamefont {J.~J.}\ \bibnamefont {Relancio}},\ and\ \bibinfo {author} {\bibfnamefont {M.~A.}\ \bibnamefont {Reyes}},\ }\bibfield  {title} {\bibinfo {title} {{Decay of superluminal neutrinos in the collinear approximation}},\ }\href {https://doi.org/10.1103/PhysRevD.107.043001} {\bibfield  {journal} {\bibinfo  {journal} {Phys. Rev. D}\ }\textbf {\bibinfo {volume} {107}},\ \bibinfo {pages} {043001} (\bibinfo {year} {2023})},\ \Eprint {https://arxiv.org/abs/2210.02222} {arXiv:2210.02222 [hep-ph]} \BibitemShut {NoStop}%
\bibitem [{\citenamefont {Cohen}\ and\ \citenamefont {Glashow}(2011)}]{Cohen:2011hx}%
  \BibitemOpen
  \bibfield  {author} {\bibinfo {author} {\bibfnamefont {A.~G.}\ \bibnamefont {Cohen}}\ and\ \bibinfo {author} {\bibfnamefont {S.~L.}\ \bibnamefont {Glashow}},\ }\bibfield  {title} {\bibinfo {title} {{Pair Creation Constrains Superluminal Neutrino Propagation}},\ }\href {https://doi.org/10.1103/PhysRevLett.107.181803} {\bibfield  {journal} {\bibinfo  {journal} {Phys. Rev. Lett.}\ }\textbf {\bibinfo {volume} {107}},\ \bibinfo {pages} {181803} (\bibinfo {year} {2011})},\ \Eprint {https://arxiv.org/abs/1109.6562} {arXiv:1109.6562 [hep-ph]} \BibitemShut {NoStop}%
\bibitem [{\citenamefont {Huo}\ \emph {et~al.}(2012)\citenamefont {Huo}, \citenamefont {Li}, \citenamefont {Liao}, \citenamefont {Nanopoulos},\ and\ \citenamefont {Qi}}]{Huo:2011ve}%
  \BibitemOpen
  \bibfield  {author} {\bibinfo {author} {\bibfnamefont {Y.}~\bibnamefont {Huo}}, \bibinfo {author} {\bibfnamefont {T.}~\bibnamefont {Li}}, \bibinfo {author} {\bibfnamefont {Y.}~\bibnamefont {Liao}}, \bibinfo {author} {\bibfnamefont {D.~V.}\ \bibnamefont {Nanopoulos}},\ and\ \bibinfo {author} {\bibfnamefont {Y.}~\bibnamefont {Qi}},\ }\bibfield  {title} {\bibinfo {title} {{Constraints on Neutrino Velocities Revisited}},\ }\href {https://doi.org/10.1103/PhysRevD.85.034022} {\bibfield  {journal} {\bibinfo  {journal} {Phys. Rev. D}\ }\textbf {\bibinfo {volume} {85}},\ \bibinfo {pages} {034022} (\bibinfo {year} {2012})},\ \Eprint {https://arxiv.org/abs/1112.0264} {arXiv:1112.0264 [hep-ph]} \BibitemShut {NoStop}%
\bibitem [{\citenamefont {Kostelecky}\ and\ \citenamefont {Russell}(2011)}]{Kostelecky:2008ts}%
  \BibitemOpen
  \bibfield  {author} {\bibinfo {author} {\bibfnamefont {V.~A.}\ \bibnamefont {Kostelecky}}\ and\ \bibinfo {author} {\bibfnamefont {N.}~\bibnamefont {Russell}},\ }\bibfield  {title} {\bibinfo {title} {{Data Tables for Lorentz and CPT Violation}},\ }\href {https://doi.org/10.1103/RevModPhys.83.11} {\bibfield  {journal} {\bibinfo  {journal} {Rev. Mod. Phys.}\ }\textbf {\bibinfo {volume} {83}},\ \bibinfo {pages} {11} (\bibinfo {year} {2011})},\ \Eprint {https://arxiv.org/abs/0801.0287} {arXiv:0801.0287 [hep-ph]} \BibitemShut {NoStop}%
\bibitem [{\citenamefont {Kuznetsov}(2024)}]{Kuznetsov:2023jfw}%
  \BibitemOpen
  \bibfield  {author} {\bibinfo {author} {\bibfnamefont {M.~Y.}\ \bibnamefont {Kuznetsov}},\ }\bibfield  {title} {\bibinfo {title} {{A nearby source of ultra-high energy cosmic rays}},\ }\href {https://doi.org/10.1088/1475-7516/2024/04/042} {\bibfield  {journal} {\bibinfo  {journal} {JCAP}\ }\textbf {\bibinfo {volume} {04}},\ \bibinfo {pages} {042}},\ \Eprint {https://arxiv.org/abs/2311.14628} {arXiv:2311.14628 [astro-ph.HE]} \BibitemShut {NoStop}%
\bibitem [{\citenamefont {Adriani}\ \emph {et~al.}(2025{\natexlab{c}})\citenamefont {Adriani} \emph {et~al.}}]{KM3NeT:2025mfl}%
  \BibitemOpen
  \bibfield  {author} {\bibinfo {author} {\bibfnamefont {O.}~\bibnamefont {Adriani}} \emph {et~al.} (\bibinfo {collaboration} {KM3NeT}),\ }\bibfield  {title} {\bibinfo {title} {{KM3NeT Constraint on Lorentz-Violating Superluminal Neutrino Velocity}},\ }\href@noop {} {\  (\bibinfo {year} {2025}{\natexlab{c}})},\ \Eprint {https://arxiv.org/abs/2502.12070} {arXiv:2502.12070 [astro-ph.HE]} \BibitemShut {NoStop}%
\end{thebibliography}%
\end{document}